\documentclass[11pt,fleqn]{article}
\usepackage{psfig}
\title{Possible Zero-Flux Transport induced by \\ Density Waves  in a  Tube filled with Solid
Helium}
\author{Kwang-Hua W. Chu \cite{2007:CHU}}  %affiliation
\date{Department of Physics, Xinjiang University, Urumqi 830046, PR China}
\begin{document}
\maketitle                 % Produces the title.
%%\documentstyle[11pt,leqno,fleqn]{article}     %\parindent=0 pt
%%\renewcommand{\baselinestretch}{1.5}
%%\evensidemargin-12mm%%\renewcommand{\baselinestretch}{1}
%%\baselinestretch=1        %\intextskip=6mm
\begin{abstract}
%\doublerulesep=8mm        %\parskip=12 pt
%\baselineskip=8mm
Macroscopic derivation of the entrainment in  a supersolid cylinder
 induced by a surface elastic wave (of small amplitude) propagating along the
flexible interface is conducted by considering the nonlinear
coupling between the interface and the rarefaction effect. We obtain
the critical bounds for  zero-volume-flow-rate states corresponding
to specific rarefaction measure and wave number which is relevant to
the rather small critical velocity or disappearance of supersolid
flows reported by  Rittner and  Reppy.\newline

\end{abstract}
%\pacs{67.80.Cx; 67.90.+z; 68.49.Bc; 34.50.Dy; }
\doublerulesep=6.5mm        %\parskip=12 pt
\baselineskip=6.5mm         %
\oddsidemargin-1mm
%\vspace{4mm}
%\evensidemargin+4mm
%----------------------------------------------------------------------
\bibliographystyle{plain}
\section{Introduction}
In 1969, it was conjectured by Andreev and Lifshitz$^{1}$
that at zero temperature, delocalized defects may exist
in a quantum solid, as a result of which the number of
sites of an ideal crystal lattice may not coincide with the
total number of particles. Originally, this conjecture was
proposed for three dimensional quantum solids made of
atoms ($^3$He, $^4$He, $\cdots$) which do not interact via Coulomb
repulsion.
%Day {\it et al.} [1] recently have  used a capacitive technique to directly monitor density changes
%for helium confined in Vycor at low temperature and have used a piezoelectrically driven diaphragm
%to study the pressure-induced flow of solid helium into the Vycor pores. Their measurements showed
%no indication of a mass redistribution in the Vycor that could mimic supersolid [2-8] decoupling and thus they put
%an upper limit of about 0.003 $\mu$m/s on any pressure-induced supersolid flow in the pores of Vycor.
The proposed {\it supersolid} phase is believed to occur due to the quantum
behavior of point defects, namely vacancies and interstitials, in this
crystal of bosons$^{2-3}$.
%Day {\it et al.} [1] finally claimed that, if a supersolid exists, then
%its flow properties must be quite different from that of
%superfluids, since the chemical potential difference created by a pressure change does not appear to produce
%superflow. As commented in [5], Beamish have conducted a study [9] looking at the density variations of solid
%$^4$He between two capacitor plates and it did not reveal any
%presence of vacancies.
Researchers have found that
a small lattice model does not exhibit the mesoscopic signature
of an intermediate phase separating the solid from
the liquid, where the solid and the fluid would coexist$^{4}$.
Such a vacancy-solid phase was indeed suggested$^{1}$ by
Andreev and Lifshitz if the zero point motions of certain
defects become sufficient to form waves propagating inside
the solid.  \newline
Castaing and Nozi\`{e}res have later considered$^{5}$
such a possibility for spin polarized $^3$He. The statistics of
the defects depend on their nature.
For simple vacancies
in the crystal, their statistics is given by the statistics of
the particles out of which the solid is made. If the defects
are bosons, they may form a condensate, giving rise to a
superfluid coexisting with the solid. This supersolid phase
is discussed in certain bosonic models$^{6}$. If the defects
are fermions, they may form a Fermi liquid$^{7}$ coexisting
with the solid, such that the system is neither a solid,
nor a liquid.
Two kinds of motion are possible in it; one
possesses the properties of motion in an elastic solid, the
second possesses the properties of motion in a liquid.
This interesting issue motivates our present study.  \newline
Early theoretical work by Andreev and Lifshitz$^{1}$ and Chester$^{2}$
showed that solids may feature a Bose-Einstein condensate
of vacancies (or interstitial atoms) and thus possess
superfluid (SF) properties. Quite recently one description of the quantum
solid is as a density wave that has formed in the
quantum fluid$^{8-9}$. The periodicity of this density wave need not match
precisely to the particle density, so that the ground state may be
incommensurate, with unequal densities of
vacancies and interstitials. Whether or not the
same is true for quantum {\it fluctuations} is not clear at this point.
We noticed that previous theories imply a corresponding
vacancy contribution to the specific heat that is as large as the
phonon contribution near
1 Kelvin$^{3,10}$.
%Meanwhile, by excluding a bulk supersolid interpretation
%of the Kim and Chan results Prokof'ev and Svistunovwe have
%looked for an alternative explanation of their data based on
%the physics of disordered and frustrated $^4$He interfaces.
Based on these considerations or phenomenological approaches,
assuming the existance of small-amplitude density waves along the
deformable boundaries, in this letter, we shall demonstrate that
wavy flexible interface (between atoms and free vacancies or
defects) or highly-pressured environments$^{8}$ can produce
elastically deformed interface or peristaltic motion will induce
time-averaged transport in a Andreev-Lifshitz supersolid$^{1,2}$.
\newline
%------------------------------------------------------
Theoretical studies of interphase nonlocal transport phenomena which
appear as a result of a different type of nonequilibrium
representing propagation of a surface elastic wave have been
performed before$^{11-12}$. These are relevant to particles  flowing
along deformable elastic slabs with the dominated parameter being
the Knudsen number (Kn = mean-free-path/$L_d$, mean-free-path (mfp)
is the mean free path (of the particles) which is thus temperature
dependent$^{13,14}$, $L_d$ is proportional to the characteristic
distance between two boundaryies)$^{13-15}$. The role of the Knudsen
number is similar to that of the Navier slip$^{15}$ parameter $N_s
(= \mu S/L_d$;
 S is a proportionality constant as $u_s = S \tau$,
$\tau$ :
the shear stress of the bulk velocity; $u_s$ : the dimensional slip velocity;
for a
no-slip case, $S = 0$, but for a no-stress condition. $S=\infty$, $\mu$ is the
 viscosity). \newline
We shall choose a periodic domain
% (say, a slab region
%with a upstream inlet and a downstream outlet along one direction,
%the upper and lower interfaces
%corresponding to the   boundaries of the elastic
%slab, the other spanwise boundaries being infinitely longer than
%the length scales prescribed for two other directions)
to simplify our mathematical treatments. The deformable (elastic)
interface is presumed. We adopt the macroscopic or hydrodynamical
approach and simplify the original system of equations (related to
the momentum and mass transport) to one single higher-order
quasi-linear partial differential equation in terms of the unknown
stream function. In this study, as the temperature is rather low and
the phase is related to the supersolid (there might be weakly
friction or shearing dissipation in-between) we shall assume that
 the governing
equations are the incompressible Navier-Stokes equations which will
be associated with the microscopically slip velocity boundary
conditions
along the interfaces$^{13-15}$ (cf. Ref. 13 for the quantum slip case). %We then introduce the perturbation
%technique so that we can solve the related boundary value problem
%approximately.
To consider the originally quiescent environment for
simplicity, due to the difficulty in solving a fourth-order
quasi-linear complex ordinary differential equation (when the wavy
boundary condition are imposed), we can finally get an
analytically perturbed solution and calculate those physical
quantities, like, time-averaged transport or entrainment,
 critical  forcing corresponding to
the freezed or zero-volume-flow-rate states. The latter might be
relevant to those reported in Refs. [16-18] for very low flow
rates$^{16}$ or disappearance of supersolidity through an annealing
of the solid helium sample$^{17}$. Note that, as also reported in
Ref. 18, their results suggested that grain boundaries (GBs) are
superfluid, so that $^4$He crystals of medium quality are supersolid
at the liquid/solid equilibrium pressure (P$_m$); that is, mass
transport through them without dissipation is possible.
\section{Formulations}
To escape from the difficulties in treating many-body or
many-particle problems together with the scattering with a elastic
boundary, we adopt the hydrodynamical approach but use the
microscopic quantum slip boundary condition which takes into account
the mean free path$^{13}$ of the dilute molecular gas corresponding
to the nonzero slip velocity$^{13}$ along the interface-wall. We
consider a circular cylindrical tube (that of the vortex core) of
uniform radius filled with a homogeneous rarefied gas (Newtonian
viscous fluid). The wall of the tube (or the interface between the
inner vortex core and the outside part of it) is not absolutely
rigid, on which is imposed axisymmetric travelling sinusoidal waves
of moderate amplitude $a$ ($z$ is the axial coordinate in the wave
propagation direction). The radial displacement from the mean
position of the wall or the interface ($r=r_w$) is thus presumed to
be $\eta$, where $\eta=a \cos [2\pi (z-ct)/\lambda]$, $\lambda$ is
the wave length, and $c$ the wave speed (cf. Fig. 1). Axisymmetric
motion is assumed with $r$ measured in the direction normal to the
mean position of the wall. $u$, $v$ are the velocity components in
the $z$- and $r$-directions, respectively. \newline
\setlength{\unitlength}{1.00mm}
\begin{picture}(150,70)(0,-15)
%%\put(112,20){\circle{15}}   %%\put(133.2,20){\circle{23.5}}
%%\put(136,20){\circle{15}}
%%\put(112,20){\vector(1,0){5}} %%\put(133.2,20){\vector(-1,0){5}}
%%\put(136,20){\vector(-1,0){5}}
\put(116,22){\makebox(0,0){z}} \put(63,42){\makebox(0,0){r}}
\put(12,25){\makebox(0,0){$r_w$}}
\put(43,36){\makebox(0,0){$a=\epsilon \,r_w$}}
\put(36,6){\makebox(0,0){$\lambda$}} \put(86,3){\makebox(0,0){wave
speed $c$}} \put(85,36){\makebox(0,0){$\eta (z,t)=a \cos
\frac{2\pi}{\lambda} (z-ct)$}} \thinlines
\put(10,30){\line(1,0){100}} \put(10,10){\line(1,0){100}}
\put(60.1,35){\vector(0,1){8}} \put(100,20){\vector(1,0){15}}
\put(44,33){\vector(0,-1){3}} \put(44,25){\vector(0,1){3}}
\put(12,23){\vector(0,-1){3}} \put(12,27){\vector(0,1){3}}
\put(20,9){\line(0,-1){4}} \put(52,9){\line(0,-1){4}}
\put(26,6){\vector(-1,0){6}} \put(46,6){\vector(1,0){6}}
\put(82,6){\vector(1,0){8}} \thicklines
{\bezier{60}(10,20)(60,20)(100,20)}
{\bezier{30}(60,4)(60,19)(60,34)}
%% -----------1999 Sep. 21---------------1999 Juli 31----------- CHU, K.-H.
{\bezier{20}(20,30)(24,32)(28,32)}
{\bezier{20}(28,32)(32,32)(36,30)} {\bezier{20}(20,10)(24,8)(28,8)}
{\bezier{20}(28,8)(32,8)(36,10)} {\bezier{20}(36,30)(40,28)(44,28)}
{\bezier{20}(44,28)(48,28)(52,30)}
{\bezier{20}(36,10)(40,12)(44,12)}
{\bezier{20}(44,12)(48,12)(52,10)}
{\bezier{20}(52,30)(56,32)(60,32)}
{\bezier{20}(60,32)(64,32)(68,30)} {\bezier{20}(52,10)(56,8)(60,8)}
{\bezier{20}(60,8)(64,8)(68,10)} {\bezier{20}(68,30)(72,28)(76,28)}
{\bezier{20}(76,28)(80,28)(84,30)}
{\bezier{20}(68,10)(72,12)(76,12)}
{\bezier{20}(76,12)(80,12)(84,10)}
{\bezier{20}(84,30)(88,32)(92,32)}
{\bezier{20}(92,32)(96,32)(100,30)} {\bezier{20}(84,10)(88,8)(92,8)}
{\bezier{20}(92,8)(96,8)(100,10)}
%%{\bezier{25}(112,32)(124,32)(124,20)}
%%{\bezier{25}(112,8)(124,8)(124,20)}
%%{\bezier{25}(136,32)(124,32)(124,20)}
%%{\bezier{25}(136,8)(124,8)(124,20)}
%%{\bezier{25}(136,32)(148,32)(148,20)}
%%{\bezier{25}(136,8)(148,8)(148,20)}
%%%\put(112,20.1){\oval(21,21)[br]}
%%\put(112,19.9){\oval(21,21)[tr]}
%%\put(112,20.1){\oval(21,21)[bl]}
%%\put(112,19.9){\oval(21,21)[tl]}
%%\put(133.2,20.1){\oval(21,21)[br]}
%%\put(133.2,19.9){\oval(21,21)[tr]}
%%\put(133.2,20.1){\oval(21,21)[bl]}
%%\put(133.2,19.9){\oval(21,21)[tl]}
%%\put(110,20){\circle{60.5}}
%%\put(140,20){\circle{60.5}}
%%\put(21.6,22.6){\makebox(0,0){$\theta+\pi/2$}}
\put(2,-10){\makebox(0,0)[bl]{\small Fig. 1\hspace*{2mm} Schematic
diagram of the wavy motion of the interface or tube wall. $\epsilon
\ll 1$}}
\end{picture}

\noindent Firstly, it is necessary to simplify these equations by
introducing nondimensional variables. We have a characteristic
velocity $c$ and three characteristic lengths $a$, $\lambda$, and
$r_w$. Thus, the following variables based on $c$ and $r_w$ could be
introduced :
\begin{displaymath}
 r'=\frac{r}{r_w}, \hspace*{2mm} z'=\frac{z}{r_w}, \hspace*{2mm}
 u'=\frac{u}{c}, \hspace*{2mm} v'=\frac{v}{c}, \hspace*{2mm}
%\end{displaymath}
%\begin{displaymath}
 \eta'=\frac{\eta}{r_w}, \hspace*{2mm} \psi'=\frac{\psi}{c\,r_w^2}, \hspace*{2mm}
 t'=\frac{c\,t}{r_w}, \hspace*{2mm} p'=\frac{p}{\rho c^2},
\end{displaymath}
where $\psi$ is the stream function, $\rho, p$ are the density and
pressure of the fluid. The amplitude ratio $\epsilon$ (presumed to
be rather small), the wave number $\alpha$, and the Reynolds number
$Re$ are defined by
\begin{displaymath}
 \epsilon=\frac{a}{r_w}, \hspace*{4mm} \alpha=\frac{2 \pi r_w}{\lambda},
 \hspace*{4mm} Re =\frac{c\,r_w}{\nu}.
\end{displaymath}
We shall seek a solution in the form of a series in the parameter
$\epsilon$ :
\begin{displaymath}
 \psi=\psi_0 +\epsilon \psi_1 + \epsilon^2 \psi_2 + \cdots,
 \hspace*{12mm} \frac{\partial p}{\partial z}=(\frac{\partial p}{\partial
 z})_0+\epsilon (\frac{\partial p}{\partial z})_1 +\epsilon^2 (\frac{\partial
 p}{\partial z})_2 +\cdots,
\end{displaymath}
with $v=(\partial \psi/\partial z)/r$, $u=-(\partial \psi/\partial
r)/r$. \newline The $r$- and $z$-momentum equations and the equation
of continuity$^{19,20}$ could be in terms of the stream function
$\psi$ if the pressure ($p$) term is eliminated. The final governing
equation is
\begin{equation}
 \frac{\partial}{\partial t} \hat{\nabla}^2 \psi + \frac{\psi_z}{r}
 [\hat{\nabla}^2
 \psi_r-\frac{2}{r} \hat{\nabla}^2 \psi +\frac{\psi_r}{r^2}]
 -\frac{\psi_r}{r} \hat{\nabla}^2 \psi_z =\frac{1}{Re}\hat{\nabla}^4 \psi,
\hspace*{6mm} \hat{\nabla}^2 \equiv\frac{\partial^2}{\partial z^2}
+\frac{\partial^2}{\partial r^2}-\frac{1}{r}\frac{\partial}{\partial
r},
\end{equation}
and subscripts indicate the partial differentiation. Thus, we have
\begin{equation}
 \frac{\partial}{\partial t} \hat{\nabla}^2 \psi_0 +
 \frac{\psi_{0_z}}{r} [\hat{\nabla}^2
 \psi_{0_r}-\frac{2}{r} \hat{\nabla}^2 \psi_0 +\frac{\psi_{0_r}}{r^2}]
 -\frac{\psi_{0_r}}{r} \hat{\nabla}^2 \psi_{0_z}
 =\frac{1}{Re}\hat{\nabla}^4 \psi_0,
\end{equation}
\begin{displaymath}
 \frac{\partial}{\partial t} \hat{\nabla}^2 \psi_1 +
 \frac{\psi_{1_z}}{r} [\hat{\nabla}^2
 \psi_{0_r}-\frac{2}{r} \hat{\nabla}^2 \psi_0 +\frac{\psi_{0_r}}{r^2}]
 -\frac{\psi_{1_r}}{r} \hat{\nabla}^2 \psi_{0_z} +
\end{displaymath}
\begin{equation}
 \hspace*{12mm}  \frac{\psi_{0_z}}{r} [\hat{\nabla}^2
 \psi_{1_r}-\frac{2}{r} \hat{\nabla}^2 \psi_1 +\frac{\psi_{1_r}}{r^2}]
 -\frac{\psi_{0_r}}{r} \hat{\nabla}^2
 \psi_{1_z}=\frac{1}{Re}\hat{\nabla}^4 \psi_1,
\end{equation}
\begin{displaymath}
 \frac{\partial}{\partial t} \hat{\nabla}^2 \psi_2 + \frac{\psi_{0_z}}{r}
[\hat{\nabla}^2
 \psi_{2_r}-\frac{2}{r} \hat{\nabla}^2 \psi_2 +\frac{\psi_{2_r}}{r^2}]
 -\frac{\psi_{0_r}}{r} \hat{\nabla}^2 \psi_{2_z} +\frac{\psi_{2_z}}{r}
 [\hat{\nabla}^2
 \psi_{0_r}-\frac{2}{r} \hat{\nabla}^2 \psi_0 +\frac{\psi_{0_r}}{r^2}]
\end{displaymath}
\begin{equation}
 \hspace*{6mm} -\frac{\psi_{2_r}}{r} \hat{\nabla}^2 \psi_{0_z}+\frac{\psi_{1_z}}{r}
 [\hat{\nabla}^2
 \psi_{1_r}-\frac{2}{r} \hat{\nabla}^2 \psi_1 +\frac{\psi_{1_r}}{r^2}]
 -\frac{\psi_{1_r}}{r} \hat{\nabla}^2 \psi_{1_z}
 =\frac{1}{Re}\hat{\nabla}^4 \psi_2,
\end{equation}
and other higher order forms. The fluid is subjected to boundary
conditions imposed by the symmetric motion of the walls and the
non-zero slip velocity$^{13,14}$ : $u=$ -Kn $du/dr$, $v=
\partial \eta/\partial t$ at $r=(1+ \eta)$. The boundary
conditions may be expanded in powers of $\eta$ and then $\epsilon$ :
\begin{displaymath}
 [(\frac{-1}{r}\frac{\partial}{\partial r})(\psi_0 +\epsilon\psi_1 +\epsilon^2
 \psi_2+\cdots)]|_{r=1+\epsilon \cos \alpha (z-t)} =- \mbox{Kn}
 [(\frac{1}{r^2}\frac{\partial}{\partial r}-\frac{\partial^2}{r\partial
 r^2}) (\psi_0 +
\end{displaymath}
\begin{equation}
 \hspace*{24mm} \epsilon \psi_1 +\epsilon^2 \psi_2+\cdots)]|_{r=1+\epsilon
 \cos \alpha (z-t)} ,
\end{equation}
%\begin{equation}
%\end{equation}
\begin{displaymath}
 \psi_{0z}|_1 +\epsilon [\cos \alpha (z-t) \psi_{0zr}|_1 +\psi_{1z}|_1
 ]+\epsilon^2 [\frac{\psi_{0zrr}|_1}{2} \cos^2 \alpha (z-t)+\cos \alpha
 (z-t) \psi_{1zr}|_1 +
\end{displaymath}
\begin{equation}
 \hspace*{25mm} \psi_{2z}|_1 ] +\cdots =\epsilon \alpha \sin \alpha (z-t)+\epsilon^2
 \alpha \cos \alpha (z-t) \sin \alpha (z-t) .
\end{equation}
Equations above, together with the condition of symmetry and a
uniform pressure-gradient in the z-direction, $(\partial p/\partial
z)_0$=constant, yield :
\begin{equation}
 \psi_0 =K_0 [r^2-\frac{r^4}{2}],  \hspace*{24mm}
 K_0=\frac{Re}{8}(\frac{\partial p}{\partial z})_0 ,
\end{equation}
\begin{equation}
 \psi_1 = \phi(r) e^{i \alpha (z-t)}+\phi^* (r) e^{-i \alpha
 (z-t)}+\phi_0 (r) ,
\end{equation}
where the asterisk denotes the complex conjugate. A substitution of
$\psi_1$ into equation (3) yields
\begin{equation}
 \{\frac{d^2}{d r^2} -\alpha^2 -\frac{d}{r\,d r}+i \alpha Re [1+2 K_0 (1-r^2)]\}
 (\frac{d^2}{d r^2} -\alpha^2 -\frac{d}{r\,d r}) \phi =0, \hspace*{6mm}
 \hat{\nabla}^2 \hat{\nabla}^2 \phi_0 =0.
\end{equation}
The boundary conditions are
\begin{equation}
 \phi_r|_1 = 2K_0 , \hspace*{12mm}  {\phi_0}_r |_1 =4 K_0 \mbox{Kn},
 \hspace*{12mm}  \phi (1)=-\frac{1}{2}.
\end{equation}
The equations for $\phi^*$ are conjugate to the foregoing and need
not be written down. \newline Similarly, with
\begin{equation}
  \psi_2=D(r)+E(r) e^{i 2\alpha (z-t)} +E^* (r) e^{-i
  2\alpha (z-t)}+ G(r) e^{i \alpha (z-t)} +G^* (r) e^{-i
  \alpha (z-t)} ,
\end{equation}
we have
\begin{displaymath}
  (\frac{d^2}{d r^2} -\frac{d}{r d r}) (\frac{d^2}{d r^2} -\frac{d}{r d r})
  D=i \alpha Re\frac{d}{dr }\{\frac{1}{r}[(\phi \phi^*_{rr}-\phi_{rr} \phi^*)-\frac{1}{r}
  (\phi \phi^*_r -\phi_r \phi^*)] \} - \frac{i \alpha Re}{r^2}
\end{displaymath}
\begin{equation}
 \hspace*{6mm}[(\phi \phi^*_{rr}-\phi_{rr} \phi^*)-\frac{1}{r}
  (\phi \phi^*_r -\phi_r \phi^*)] \equiv i\alpha Re \{\frac{d}{dr}[\frac{(S_1-r^{-1} S_2)}{r}]
  -\frac{(S_1-r^{-1} S_2 )}{r^2} \}.
\end{equation}
\begin{displaymath}
 (\frac{d^2}{d r^2} -4\alpha^2 -\frac{d}{r dr})^2 E=-i 2 \alpha Re [1+ 2K_0 (1-r^2)]
  (\frac{d^2}{d r^2} - 4\alpha^2 -\frac{d}{r dr}) E +
\end{displaymath}
\begin{equation}
 \hspace*{12mm} \frac{i \alpha Re}{r}[\phi\phi_{rrr}-\phi_r
 \phi_{rr} -\frac{3}{r} \phi \phi_{rr} +\frac{1}{r}
 (\phi_r)^2+\frac{3}{r} \phi \phi_r +\frac{2}{r} \alpha^2 \phi^2];
\end{equation}
\begin{displaymath}   %(1+2 K_n)
 \{\frac{d^2}{d r^2} -\alpha^2 -\frac{d}{r dr}+i \alpha Re [1+2 K_0 (1-r^2)]\}
 (\frac{d^2}{d r^2} -\alpha^2-\frac{d}{r dr}) G = \frac{ i \alpha Re}{r}
 [\phi {\phi_0}_{rrr} -
\end{displaymath}
\begin{equation}
 \hspace*{12mm} 3\frac{\phi {\phi_0}_{rr}}{r}+3\frac{\phi
 {\phi_0}_r}{r^2} +\alpha^2 \phi {\phi_0}_r -\phi_{rr} {\phi_0}_r
 +\frac{\phi {\phi_0}_r}{r}].
\end{equation}
and the boundary conditions
\begin{displaymath}
 D_r (1) +\frac{1}{2} [\phi_{rr} (1)+\phi^*_{rr} (1)] -3
 K_0= 0,
\end{displaymath}
%\begin{equation}
%  \hspace*{24mm} \phi_r (1)+\phi^*_r (1)-\phi_{rr} (1)-\phi^*_{rr} (1)] +2 K_0 \},
%\end{equation}
\begin{equation}
 E_r (1)+\frac{1}{2} \phi_{rr} (1) -\frac{3 K_0}{2} = 0 ,
 \hspace*{12mm}
%\end{equation}
%\begin{equation}
 E(1)+\frac{1}{4} \phi_r (1) =-\frac{1}{4} ;
\end{equation}
\begin{equation}
 G_r (1)+\frac{{\phi_0}_{rr}(1)}{2}=\mbox{Kn} [4 K_0
 -\phi_{rr}(1)+\frac{\phi_r (1)}{r}],  \hspace*{12mm} G(1)=\frac{\phi^*_r (1)-\phi_r (1)}{2}=0.
\end{equation}
Here, $S_1 \equiv S_1(r)=\phi \phi^*_{rr}-\phi_{rr} \phi^*$, $S_2
\equiv S_2 (r)=\phi \phi^*_r -\phi_r \phi^*$. The equations for
$E^*$ are conjugate to those for $E$. We can use these equations to
determine the solution up to $O(\epsilon^2)$. The boundary
conditions are not linearized. Equations (14-16) account for the
effect of satisfying the velocity-slip condition at the wavy
interface, rather at the mean position of the interface.
\newline To illustrate the nature of the solution, particularly
with respect to the effects of nonlinearities in convective
acceleration and boundary conditions, we shall consider the
important case of pumping in the absence of the zeroth-order
pressure-gradient; i.e., when $(\partial p/\partial z)_0 =0$.
However, a simple superposition of a pressure-driven flow
corresponding to equation (7) with $K_0 \sim O(\epsilon^2)$ is
permissible.
\subsection{Originally Quiescent Environment}
To simplify the approach and obtain preliminary analytical solutions
of above complicated equations and boundary conditions, we only
consider the case in which $(\partial p/\partial z)_0$ vanishes or
$K_0=\psi_0=0$. Kn is presumed to be comparable with the order of
magnitude $O(\epsilon)$ and will be rescaled firstly (like
$\hat{\mbox{Kn}}$) and then adopted by the same representation in
the following. Hence equations (9-10) become
\begin{equation}
 (\frac{d^2}{d r^2} -\frac{d}{d r}-\alpha^2) (\frac{d^2}{d r^2}-\frac{d}{d r} -\bar{\alpha}^2)
 \phi =0 ,   \hspace*{24mm} \bar{\alpha}^2= \alpha^2 -i \alpha Re ,
\end{equation}
\begin{equation}
  \phi_r (1)=0, \hspace*{24mm} \phi (1)=-\frac{1}{2};
\end{equation}
\begin{equation}
 \hat{\nabla}^2 \hat{\nabla}^2 \phi_0 =0, \hspace*{24mm} {\phi_0}_r (1)=0;
\end{equation}
together with the condition that the velocity, i.e., $(\phi_r
/r)|_{r=0}$ remains finite along the axis of the tube (at $r=0$).
After lengthy algebraic manipulations, we obtain
\begin{equation}
 {\phi_0}_r = Q_0 \,r(1-r^2);  \hspace*{24mm} \phi=A r I_1 (\bar{\alpha} r)+ B r I_1
 (\alpha r),
\end{equation}
where $Q_0$ is an integration constant; $A=A_0/det$, $B=B_0/det$;
$A_0= \alpha I_0 (\alpha)/2$, $B_0= -\bar{\alpha} I_0
(\bar{\alpha})/2$; and
\begin{displaymath}
 det=\bar{\alpha} I_0 (\bar{\alpha})  I_1 (\alpha)- \alpha I_0
 (\alpha) I_1 (\bar{\alpha}) .
\end{displaymath}
%\begin{displaymath} %\end{displaymath}
Meanwhile, equation (12) becomes
\begin{equation}
 D_{rrrr} -\frac{2 D_{rrr}}{r}+\frac{3 D_{rr}}{r^2}-\frac{3
 D_r}{r^3}=-\frac{d L(r)}{dr}+\frac{L(r)}{r}=\bar{S}(r), \hspace*{4mm}
 L(r)\equiv-D_{rrr}+\frac{D_{rr}}{r}-\frac{D_r}{r^2}.
\end{equation}
where $\bar{S}(r)$ denotes the right-hand-side (RHS) term of
equation (12). Because of equations (12,20), equation (21) can be
directly solved, if we denote $\bar{S}(r) =-\alpha^2 Re^2 S(r)$, so
that we have
\begin{displaymath}
 D(r)=\int^r \frac{-\bar{S}(t) t^3 dt}{16}+\frac{r^2}{4}\int^r \bar{S}(t)\, t \, \log(t) dt+
 \frac{r^4}{16} \int^r \frac{\bar{S}(t)}{t} dt -\frac{\log(r) r^2}{4}\int^r \bar{S}(t) t dt +
\end{displaymath}
\begin{equation}
 \hspace*{24mm} b_4 +b_3 r^4 +b_2 r^2 \log(r)+ b_1 r^2 ,
\end{equation}
where $b_i, i=1,2,3,4$ are integration constants. To obtain a simple
solution which relates to the mean flow so long as only terms of
$O(\epsilon^2)$ are concerned, we see that if every term in the
z-momentum equation is averaged over an interval of time equal to
the period of oscillation [9], we obtain for our solution as given
by above equations the mean pressure gradient
\begin{displaymath}
 \overline{\frac{\partial p}{\partial z}}=\epsilon \overline{(\frac{\partial
 p}{\partial z})_1}+ \epsilon^2 \overline{(\frac{\partial
 p}{\partial z})_2}+O (\epsilon^3) =\frac{\epsilon}{r\,Re}[-{\phi_0}_{rrr}+
 \frac{{\phi_0}_{rr}}{r}-\frac{{\phi_0}_r}{r^2}]+
 \epsilon^2 \{\frac{1}{r Re} (-D_{rrr}+\frac{D_{rr}}{r}-
\end{displaymath}
\begin{equation}
 \hspace*{12mm}\frac{D_r}{r^2}) +\frac{i \alpha}{r^2} [\phi \phi^*_{rr} -\phi^* \phi_{rr}-
 \frac{1}{r}(\phi \phi^*_r- \phi^* \phi_r)] \}+ O(\epsilon^3).
\end{equation}
Thus, as far as the mean flow is concerned $D$ is the only term
which participates in the solution as long as only terms of
$O(\epsilon)$ are retained. We have no need to consider $E$ when
considering the mean free flow. Actually, with equations (12,21) and
the rearrangement of both sides, we have
\begin{equation}
 \overline{(\frac{\partial p}{\partial z})_1}=\frac{4 Q_0}{Re},
 \hspace*{24mm}
 \overline{(\frac{\partial p}{\partial z})_2}=\frac{P_0}{Re},
\end{equation}
where $P_0$ is an integration constant by considering the
recombination of equations (12,21) and the integration hereafter
(cf. equation (23)).
\newline Now, from equation (14), we have
\begin{equation}
  D_r (1)= -\frac{1}{2} [\phi_{rr} (1)+\phi^*_{rr}(1)]
\end{equation}
here, we denote $\zeta=-[\phi_{rr}(1)+\phi^*_{rr}(1)]/2$.
\newline From equation (22) and differentiating $D$ once, we
obtain
\begin{equation}
  D_r (r)= (a_1 +a_2 ) r +2 a_2 r \ln r+ a_0 r^3 +\bar{G} (r), \hspace*{12mm}
  \bar{G}(r)\equiv-\alpha^2 Re^2 G(r),
\end{equation}
where $a_2$ must be zero as the axial velocity which is proportional
to $D_r /r$ remains finite at $r=0$. Besides, from the expression of
$D(r)$, we know that $\bar{G} (r) /r=0$ at $r=0$. \newline From
equations (22-23,25), we know that $P_0=-4 a_0$. $a_0$ or $P_0$ is
then proportional to the time-averaged pressure-gradient (on the
axis) accompanying the peristaltic motion and now
\begin{equation}
 D_r (r)= [\zeta - \bar{G}(1)] r +
%\end{displaymath}
%\begin{equation}
  a_0[r^3-r]+\bar{G} (r).
\end{equation}
With $a_0$ or $P_0$ specified, the solution for the mean axial
velocity (averaged over time) is
\begin{displaymath}
 \bar{u}\equiv U(r)=-\epsilon \frac{{\phi_0}_r}{r}-\epsilon^2\frac{D_r}{r}=\epsilon [-Q_0
 (1-r^2)]+ \epsilon^2 \{-[\zeta - \bar{G}(1)]-
% {\cal C} (y)=\frac{\alpha^2 Re^2}{2} [\frac{c_0 c_2^*}{g_1^2}
% e^{(\alpha+\bar{\alpha}^*) y} + \frac{c_0^* c_2}{g_2^2} e^{(\alpha+\bar{\alpha}) y} +
% \frac{c_0 c_3^*}{g_3^2} e^{(\alpha-\bar{\alpha}^*) y} + \frac{c_0^*
% c_3}{g_4^2} e^{(\alpha-\bar{\alpha}) y} +\frac{c_1 c_2^*}{g_3^2} e^{
% (\bar{\alpha}^*-\alpha) y}
%\end{displaymath}
%\begin{equation}
 \frac{\bar{G}(r)}{r}+a_0(1-r^2)\}=
\end{displaymath}
\begin{equation}
 \hspace*{16mm} \epsilon [-Q_0 (1-r^2)]+\epsilon^2 \{-R_0+ \bar{G} (1)-
 \frac{\bar{G}(r)}{r}+a_0(1-r^2) \}.
\end{equation}
In practical applications we must determine $a_0$ from
considerations of conditions at the ends of the channel. As usual, a
critical reflux condition  can be defined as the condition where
there is zero velocity at the center of the tube. Using equations
(22-23,27), we have
\begin{equation}
 \overline{(\frac{\partial p}{\partial z})_2}|_{cr}=\frac{4}{Re}
 [-R_0+ \bar{G} (1)],
\end{equation}
where $R_0$ becomes $\zeta$ (cf. Ref. 19) as  Kn=0 and is due to the
nonlinear effect of the boundary condition.
%------------------------------------------------- Replace
\section{Results and Discussion}
%We check our approach firstly by examining $R_0$ with that of
%no-slip (Kn$=0$) approach. This can be done easily once
%we consider terms of $D_y(y)$ and ${\cal C} (y)$ because to
%evaluate $R_0$ we shall at most take into account the higher
%derivatives of $\phi(y)$, like $\phi_{yy} (y)$, $\phi_{yyy} (y)$
%instead of $\phi_y (y)$ and escape from the prescribing of $a_2$. \newline
Our numerical calculations confirm that the mean streamwise velocity
distribution (averaged over time) due to the induced motion by the
wavy elastic  interface in the case of free (vacuum) pumping is
dominated by $R_0$ (or Kn) and the parabolic distribution $-P_0
(1-r^2)$. $R_0$ which defines the boundary value of $D_r$ has its
origin in the y-gradient of the first-order streamwise velocity
distribution.
%\newline %In addition to the terms mentioned above, there is a
%perturbation term which varies across the channel : ${\cal C} (y)
%-{\cal C} (1)$. Let us define it to be
%\begin{equation}
% $F(y)={-200} [{\cal C} (y) -{\cal C} (1)]/({\alpha^2 Re^2})$.
%\label{eq22}.
%\end{equation}
%To compare with no-slip (Kn$=0$) results, we plot
%three cases, $\alpha=0.1, 0.4$, and $0.8$ for the same Reynolds number $Re=1$
% of our
%results : Kn$=0.1$ with those Kn$=0$ into Fig. 2.
Note that the Reynolds number here is based on the wave speed.  The
physical trend herein is also the same as those reported
before$^{13-14,20}$ for the slip-flow effects. The slip produces
decoupling with the inertia of the wavy interface.
\newline Now, let us define a critical reflux condition as one for
which the mean velocity ${U} (r)$ equals to zero at the center-line
$r=0$. With the equation of $U$, we have
%\begin{equation}
 $ P_{0_{cr}}=Re \overline{({\partial p}/{\partial z})_2}=
 -4 [\alpha^2
  Re^2 G(1) +R_0]$,
%-{[{\alpha^2  Re^2}F(0)/200 +\mbox{Kn} \,{\cal C}' (1)-R_0]}/{(1+2 \mbox{Kn})}$
  %\label{eq23},
%\end{equation}
which means the critical reflux condition is reached when $P_0$ has
above value. Pumping against a positive  forcing greater than the
critical value would result in a backward transport (reflux) in the
central region of the stream. This critical value depends on
$\alpha$, $Re$, and Kn. There will be no reflux if the  pressure
gradient is smaller than this $P_0$. Thus, for some $P_0$ values
less than  $P_{0_{cr}}$, the superflow will keep moving  forward. On
the contrary, parts of the flow will move
 backward
if $P_0 > P_{0_{cr}}$. \newline As reported in Refs. 8 or 17,  the
rather small critical velocity ($\le 20 \mu$m/s) observed shows an
apparent dissipation or attenuation of the superflow. Thus, we
present some of the values of $P_0(\alpha, Re; \mbox{Kn}=0, 0.15)$
corresponding to freezed or zero-volume-flow-rate states (cf. Chu in
Ref. 15) ($\int_{0}^1 U(r) dr=0$) in  Table 1 where the wave number
($\alpha$) has the range : $0.1,0.2,0.3,0.4,0.5,0.6,0.7,0.8$ and
$0.9$; the Reynolds number ($Re$)$=0.1,1,10,25, 50$ and $100$. These
freezed states might be similar to that reported in Ref. 16 (the
nonclassical rotational inertia signal (NCRI) is not a universal
property of solid $^4$He but can be eliminated through an annealing
of the solid helium sample).  \newline We observe that as Kn
increases from zero to 0.15, the critical $P_0$ decreases
significantly (cf. Fig. 2). The quantum slip parameter$^{13}$ which
is represented in terms of Kn (via the mean free path$^{13,14}$) is
temperature dependent in essence but is only relaxed in the
boundaries (weakly compressible cases here). This kind of
boundary-dominated flows resemble those reported in Ref. 18 by
Sasaki {\it et al.} For the same Kn, once Re is larger than 10,
critical reflux values $P_0$ drop rapidly and the wave-modulation
effect (due to $\alpha$) appears. The latter observation might be
interpreted as the strong coupling between the interface and the
inertia of the streaming superflow. The illustration of the velocity
fields for those zero-flux (zero-volume-flow-rate) or freezed states
are shown in Figure 3. There is one wave number : $\alpha=0.8$ and
the Reynolds number is 10. Both no-slip and slip (Kn=0.15) cases are
presented.
%The arrows for slip cases are schematic and represent the direction
%of positive and negative velocity fields.
\newline Some remarks
could be made about these states : the transport being freezed in
the time-averaged sense for specific dissipations (in terms of
Reynolds number which is the ratio of wave-inertia and viscous
shearing effects) and wave numbers (due to the wavy interface or
other fluctuations) for either no-slip and slip cases. These
resemble those reported quite recently$^{16-18}$.
 Meanwhile, the time-averaged transport induced by the wavy interface
is proportional to the square of the amplitude ratio (although the
small amplitude waves being presumed), as can be seen in Eqn. (11),
which is qualitatively the same as that presented in Ref. 11 for
analogous interfacial problems.
\newline In brief summary,
the entrained transport (either postive or negative and there is possibility : freezing)
due to the wavy  interface
 is mainly tuned
by the  $P_0$ for fixed Re. Meanwhile, $P_{0_{cr}}$ depends strongly
on the Knudsen number (Kn, a rarefaction measure) instead of Re or
$\alpha$. These results (cf. Table 1 and Fig. 2) might explain why
there are rather small critical velocities for superflows  in the
temperature range (similar to the effect represented by the quantum
slip parameter : Kn here) where a supersolidity is observed$^{8,17}$
or the disappearance of supersolidity  through an annealing of the
solid helium sample$^{16}$. We shall investigate much more
complicated problems$^{21,22}$ in the future.
%Finally we like to stress that once the phase is solid then the
%elastically tensile stress should be larger than that of the liquid
%phase and the latter could be a crucial test for the existence of
%the supersolidity.
%\newline
%\noindent     %{\large \underline{Results and Discussion}

%\psfig{file=chu-fg1.ps,bbllx=0.0cm,bblly=14.8cm,bburx=12cm,bbury=24cm,rheight=9.2cm,rwidth=9.2cm,clip=}
%
%\begin{figure} [h]
%\hspace*{3mm}
%Fig. 1 \hspace*{1mm} Schematic diagram of
%the wavy motion of the  interface.
%\end{figure}
%\begin{center}
%
\newpage
%\vspace{2mm}
%
\psfig{file=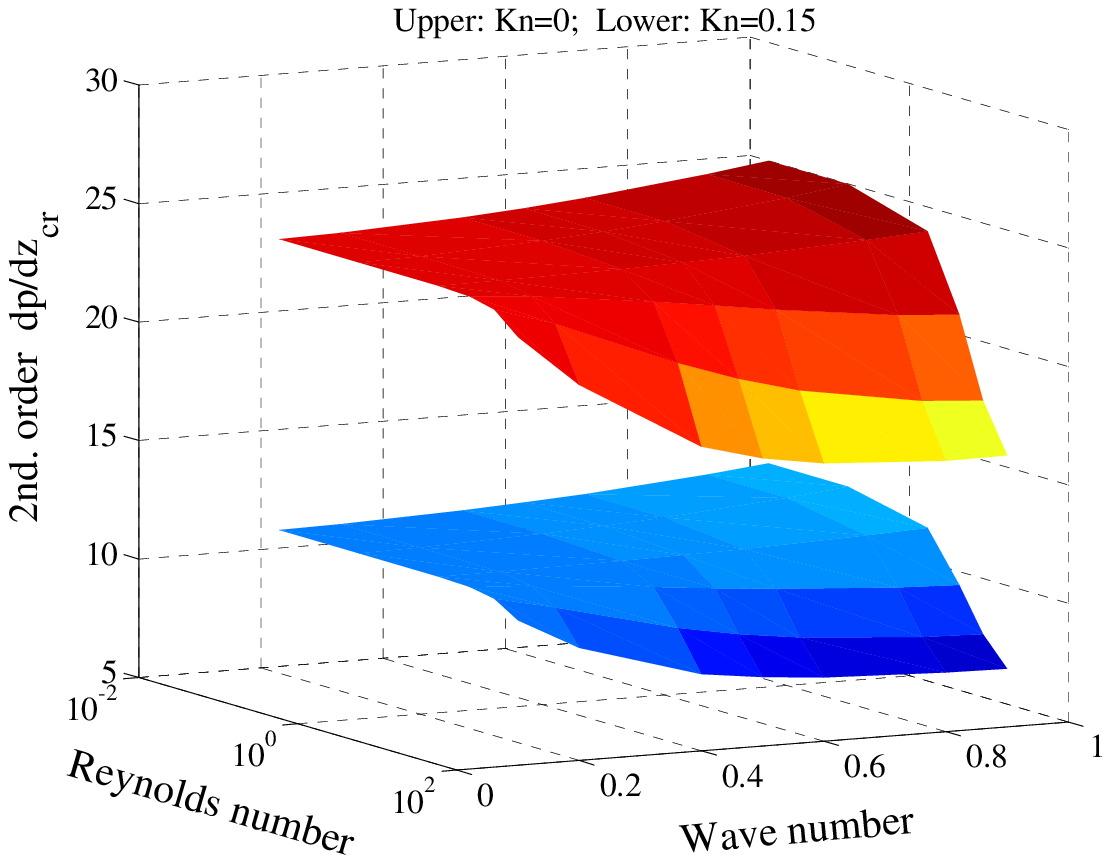,bbllx=0cm,bblly=13.8cm,bburx=12cm,bbury=24cm,rheight=10cm,rwidth=10cm,clip=}
%
%\vspace{4mm}
\begin{figure} [h]
\hspace*{3mm} Fig. 2 \hspace*{1mm} Demonstration of Kn, $Re$ and
$\alpha$ effects on the $\Pi_0$ (zero-flux states).
\newline \hspace*{3mm} %$\Pi_0 = {\Pi_0}_{cr}$.
$Re$ is the Reynolds number (the ratio of the wave-inertia and viscous
shearing dissipation).   \newline \hspace*{3mm}
$\alpha$ is the wave number and Kn is the Knudsen number
(a rarefaction measure).
\end{figure}

\newpage

\vspace{3mm}
\psfig{file=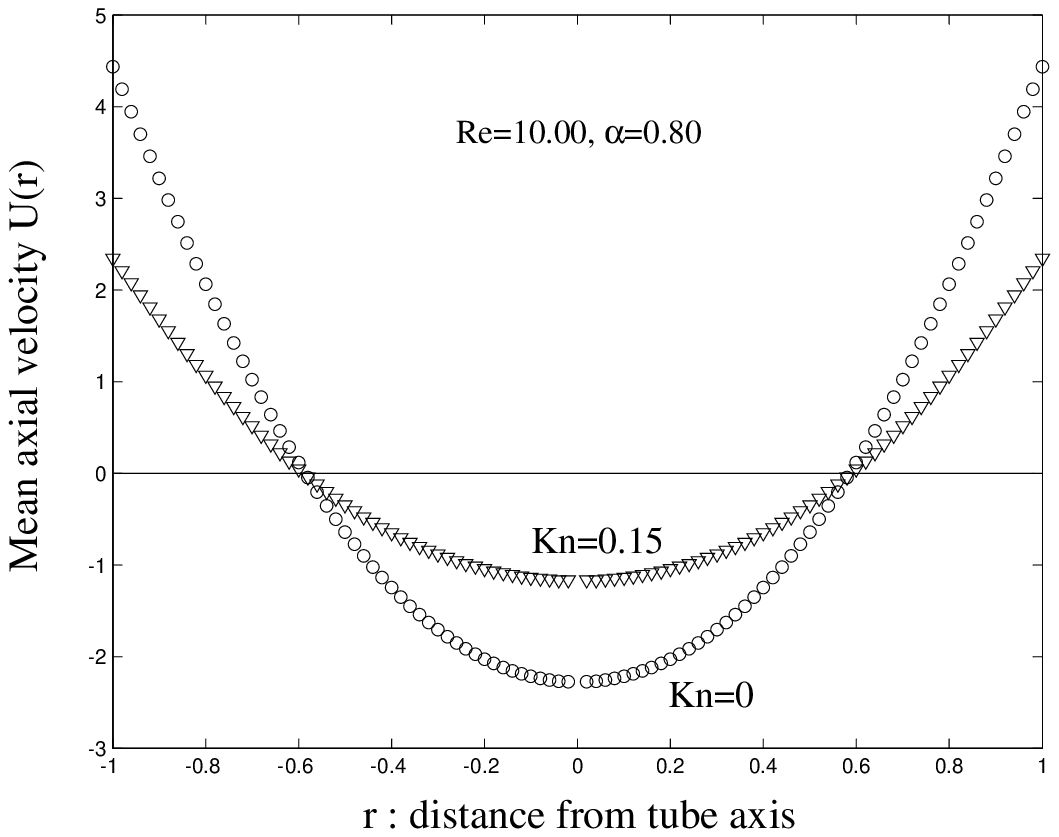,bbllx=0.0cm,bblly=12.5cm,bburx=12cm,bbury=24cm,rheight=9.8cm,rwidth=9.8cm,clip=}
%
%\end{center}
\begin{figure} [h]
%\vspace{4mm}%\vspace{4mm}
\hspace*{3mm} Fig. 3 \hspace*{1mm} Demonstration of the zero-flux
states : the mean velocity field $U(r)$ for \newline \hspace*{3mm}
wave numbers $\alpha=0.8$. The Reynolds number is $10$. Kn is the
rarefaction measure
\newline \hspace*{3mm}
(the mean free path of the particles divided by the characteristic length).
\newline \hspace*{3mm}
%The arrows are schematic and illustrate the directions of positive
%and negative $U(r)$. \newline \hspace*{3mm}
The integration of $U(r)$ w.r.t. $r$ for these velocity fields gives
zero volume flow rate.
%Upper left : $a_0 <0$; upper right : $a_0 =0$; lower left : $a_0 ={a_0}_{cr}$; lower right : $a_0 >
%{a_0}_{cr}$.  \newline \hspace*{22mm} This figure indicates the
%sequence of increasing the pressure gradient for fixed Re and $\alpha$.
\end{figure}

%\begin{figure}[h]  %\vspace{32mm} %\hspace*{25mm} Fig. (i) \hspace*{3mm}
% \caption{Schematic diagram of the wavy motion of the walls.}
 %\label{fig1}
%\end{figure}
%\begin{figure}[h]  %
%\hspace*{10mm} Fig. 2\hspace*{3mm} % \caption{
%Comparison of the mean-velocity perturbation function $F(y)$. %\newline
% %\label{fig2}
%\end{figure}
%
%\begin{figure}[h]  %
%\vspace{4mm}  %
%\hspace*{10mm} Fig. 3\hspace*{3mm} %\caption{
%Comparison of $a_0\equiv Re \overline{(dp/dx)_2}$ at the
% critical-reflux condition.
% %\label{fig3}
%\end{figure}
\newpage

\begin{table}[h]
 \caption{Zero-flux or freezed states values ($P_0$) for a cylindrical
 wavy interface.}
\vspace*{5mm}
\begin{center}
%-------------------- Tube
\begin{tabular}[b]{|r|c|c|c|c|c|c|c|}      \hline
% \multicolumn{1}{c|}{R*$\alpha$}            \hline
    &          &  Re  &   &    &    &      &      \\ \hline
 Kn & $\alpha$ &  0.1 & 1 & 10 & 25 &  50  &  100 \\ \hline
 0  & 0.1  & 24.2679 & 24.2720 &  24.2545 & 24.1846 & 23.9435 &  23.0755 \\  \cline{2-8}
    & 0.2  & 24.3293 & 24.3271 &  24.2753 & 24.0009 & 23.1347 &  20.8498  \\  \cline{2-8}
    & 0.4  & 24.5712 & 24.5731 &  24.3596 & 23.3741 & 21.0765 &  17.8457  \\  \cline{2-8}
    & 0.5  & 24.7615 & 24.7573 &  24.4307 & 23.0026 & 20.1985 &  17.1345  \\  \cline{2-8}
    & 0.6  & 24.9884 & 24.9855 &  24.5116 & 22.6346 & 19.5060 &  16.7273 \\ \cline{2-8}
    & 0.8  & 25.5830 & 25.5745 &  24.7686 & 22.0000 & 18.6721 &  16.4353 \\ \cline{2-8}
    & 0.9  & 25.9523 & 25.9414 &  24.9380 & 21.8046 & 18.4807 &  16.4546  \\ \hline
0.15 & 0.1 & 11.9953 & 11.9945 &  11.9824 & 11.9298 & 11.7435 &
11.1254 \\ \cline{2-8}
     & 0.2 & 12.0359 & 12.0359 &  11.9906 & 11.7865 & 11.1632 &   9.7550    \\ \cline{2-8}
     & 0.4 & 12.2082 & 12.2066 &  12.0421 & 11.3259 &  9.8944 &   8.2432 \\ \cline{2-8}
     & 0.5 & 12.3385 & 12.3366 &  12.0825 & 11.0743 &  9.4247 &   7.9058  \\ \cline{2-8}
     & 0.6 & 12.4980 & 12.4952 &  12.1336 & 10.8404 &  9.0827 &   7.6943  \\ \cline{2-8}
     & 0.8 & 12.9223 & 12.9152 &  12.2967 & 10.4917 &  8.7025 &   7.4861   \\ \cline{2-8}
     & 0.9 & 13.1854 & 13.1763 &  12.4088 & 10.3877 &  8.6208 &   7.4413  \\ \hline
%-------------------- Tube
\end{tabular}               %\begin{tabbing}
%%  R  \=  $\alpha$  \= mode number  \= $c_{r}$  \= $c_{i}$  \= R * $\alpha$ %\end{tabbing}
%% \caption{Certain Spectra for M=25, M is number of ploynomial-expansion terms}
\end{center}
\end{table}
\end{document}